\documentclass[10pt,aps,amssymb,floatfix,prd,twocolumn, preprintnumbers]{revtex4}
\usepackage[utf8]{inputenc}
\usepackage{epstopdf}
\usepackage{capt-of}
\usepackage{graphicx}  
\usepackage{dcolumn}   
\usepackage{bm}
\usepackage{amsmath}
\usepackage[font=scriptsize]{caption}
\usepackage[colorlinks]{hyperref}
\usepackage{xcolor}
\usepackage{orcidlink}

\begin{document}
\input epsf.tex
\title{$f(T,\mathcal{L}_m)$ Cosmology Embedded in a Viscous Barotropic Fluid}

\author{Suraj Kumar Behera \orcidlink{0009-0009-9294-460X}}
\email[]{skbehera.researches@gmail.com}
\affiliation{Department of Mathematics, School of Advanced Sciences, VIT-AP University, Beside AP Secretariat, Amaravati, 522241, Andhra Pradesh, India.}

\author{Raja Solanki}
\email[]{raja.solanki@vitap.ac.in, rajasolanki8268@gmail.com}
\affiliation{Department of Mathematics, School of Advanced Sciences, VIT-AP University, Beside AP Secretariat, Amaravati, 522241, Andhra Pradesh, India.}

\author{Pratik P. Ray \orcidlink{0000-0003-2304-0323}}
\email[]{pratik.ray@vitap.ac.in ; pratik.chika9876@gmail}
\affiliation{Department of Mathematics, School of Advanced Sciences, VIT-AP University, Beside AP Secretariat, Amaravati, 522241, Andhra Pradesh, India.}
\affiliation{Pacif Institute of Cosmology and Selfology (PICS), Sagara, Sambalpur 768224, Odisha, India}


\affiliation{}

\begin{abstract}
\begin{center}
{\textbf{Abstract}}
\end{center} 
We examine the cosmological dynamics of a viscous fluid within the framework of $f(T,\mathcal{L}_m)$ gravity by using a barotropic equation of state for cosmic fluid. The modified Friedmann equations are used to construct an analytical Hubble model. Then $H(z)$, Pantheon+SH0ES, DESI DR II BAO, and Cosmic Microwave Background datasets are used in a Bayesian Markov Chain Monte Carlo analysis to constrain its free parameters. Physically viable estimations of the Hubble constant and barotropic equation of state parameter are given by the obtained best-fit values. Based on the constrained parameters, we examine the redshift dependence of deceleration parameter, effective equation of state, dark energy equation of state, the effective pressure and viscous pressure quantities. While the deceleration parameter reveals how the universe evolved from an initially decelerating state to its present accelerating expansion, both equation of state parameters persist in the quintessence regime in the late epoch. Moreover, throughout the cosmic evolution, the viscous and effective pressures stay negative. The observational viability of the suggested viscous $f(T,\mathcal{L}_m)$ cosmological model is supported by the estimated Hubble constant, whose values agree well with the results from independent cosmological observations.
\end{abstract}

\maketitle

\section{Introduction} \label{introduction}
Understanding the physical mechanism underlying the observed dynamics of the universe is one of the fundamental goals of modern cosmology. The universe has gone through different phases of expansion, from an early radiation-dominated epoch to a matter-dominated era and finally to the present observed accelerated expansion, based on a consistent picture of cosmic evolution established over the past few decades by an extensive number of high-precision observations \cite{Tegmark_2004,Eisenstein_2005,Spergel_2003,2007ApJS..170..377S,Riess_1998,Perlmutter_1999}. Although the standard cosmological model successfully describes these stages through the introduction of dark matter (DM) and dark energy (DE) components, the actual nature of these constituents is largely unknown. Consequently, now, there is a lot of research being conducted on the hypothesis that modifications in the gravitational interaction itself could be the source of the observed cosmological dynamics. Despite the outstanding success of the $\Lambda$CDM \cite{Einstein1917} model, the investigation of other gravitational frameworks has been motivated by the unresolved problems related to DE and the cosmological constant \cite{Weinberg1989}.\\ 

One of the more intriguing theories is modified gravity (MG), which interprets the current accelerated cosmic expansion as a consequence of deviations from Einstein's GR rather than the introduction of an exotic energy component\cite{Beltran2019,Shankaranarayanan2022}. Over the years, several extensions of GR have been proposed, including $f(R)$, $f(\mathcal{G})$, scalar-tensor, teleparallel and symmetric teleparallel theories of gravity etc. These theories enrich gravitational dynamics and offer viable explanations for various phases of cosmological evolution, from inflation to the present accelerated epoch. As a result, it has become a crucial theoretical tool for comprehending the large-scale behavior of the universe. Among the different modified gravity theories, teleparallel gravity (TG) offers a different geometrical explanation of gravitation. Unlike GR, which attributes gravity to spacetime curvature, TG uses torsion scalar $T$ to explain gravitational interactions known as the teleparallel equivalent of general relativity (TEGR) \cite{golovnev2018introductionteleparallelgravities}, constructed from the tetrad fields and Weitzenb\"ock connection \cite{aldrovandi2012teleparallel}. While offering a theoretically distinct geometrical interpretation, the TEGR reproduces similar cosmological predictions as Einstein's theory at the level of the field equations. The ability of this alternative formulation to produce viable cosmological scenarios while preserving the second-order field equations has garnered a lot of interest. $f(T)$ gravity, a generalization of TG to arbitrary functions of the torsion scalar $T$, has provided new approaches for purely geometric explanations of the universe accelerated expansion. Some of the recent studies on $f(T)$ gravity are listed below. Using the full Planck datasets and late-time cosmological observations, Kumar \textit{et al.} constrained viable $f(T)$ gravity models and evaluated their consistency with recent cosmological measurements \cite{Kumar_2023}. Power-law and Linder $f(T)$ Gravity models are constrained by Escamilla-Rivera and Sandoval-Orozco using DESI BAO and DES-SN5YR data and demonstrated that the viable $f(T)$ models are still competitive with the standard $\Lambda$CDM paradigm \cite{Escamilla-Rivera:2024sae}. In power-law exponential  $f(T)$ gravity with sign-changing dark energy $z_\dagger \sim 1.5$, Akarsu \textit{et al.} discovered unexplored negative and positive - $\beta$ branches that are consistent with CMB constraints \cite{Akarsu_2025}. Golovnev and Hemida investigated the dynamical characteristics and solution space of anisotropic Bianchi-I cosmologies in $f(T)$ gravity \cite{Golovnev:2025lpt}. D'Agostino \textit{et al.} studied the influence of spatial curvature on cosmological dynamics and teleparallel dark energy models in a non-flat universe within the framework of $f(T)$ gravity \cite{Agostino_2025}. In the framework of $f(T)$ gravity, Chirde and Shekh explored the cosmological evolution of a barotropic bulk viscous fluid and discussed its associated thermodynamic properties\cite{Chirde2014}. Hashim \textit{et al.} evaluated the observational consistency of the exponential infrared $f(T)$ gravity model using recent observational datasets\cite{Hashim2026}.\\

In an effort to offer a more comprehensive explanation of gravitational phenomena, several extended teleparallel theories have been developed in response to the success of $f(T)$ gravity. Prominent extensions include $f(T,\mathcal{T})$ gravity that involves the trace $\mathcal{T}$ of the energy-momentum tensor, $f(T,B)$ gravity, where the boundary term $B$ explicitly contributes to the gravitational action, and scalar-torsion theories featuring a maximal coupling between scalar fields and torsion. These extensions enrich the gravitational dynamics and provide new insights into DE, inflation, and the late-time accelerated expansion of the universe. Among them, theories that include explicit couplings between torsion and matter have garnered special interest due to their ability  of generating novel cosmological effects beyond those of conventional $f(T)$ gravity. In this regard, $f(T,\mathcal{L}_m)$ gravity is a significant extension of TG, where the gravitational Lagrangian is expressed as function of both the torsion scalar $T$ and the matter Lagrangian density $\mathcal{L}_m$ \cite{Cruz_2026,Bahamonde_2018,Harko2014}. The gravitational field equations are modified by the explicit dependence on $\mathcal{L}_m$, which results in an exchange of information between the matter and geometrical sectors. Consequently, the ensuing cosmological dynamics can be different significantly what both conventional $f(T)$ gravity and GR predict. An interesting feature of the $f(T,\mathcal{L}_m)$ gravity is that the coupling between torsion and matter can effectively mimic DE behavior and alter the expansion history of the universe. The theory may naturally produce transitions between accelerating and decelerating phases, modify the effective equation of state (EoS), and produce distinctive signatures that can be compared to observational data, depending upon the function form selected. These characteristics make $f(T,\mathcal{L}_m)$ gravity an appealing model for investigating the recent cosmic expansion history and identifying deviations from the predictions of standard cosmology. Representative works in this direction are listed in the refs. \cite{Cruz_2026,Bahamonde_2018,Koussour_2026}.\\

In this study, assuming a barotropic equation of state (EoS) for the cosmic fluid, we explore the cosmological implications of a viscous fluid in the context of $f(T,\mathcal{L}_m)$ gravity. A Bayesian Markov Chain Monte Carlo (MCMC) algorithm is used to constrain the free model parameters using the latest cosmological observations. The best-fit values that are obtained are then used to investigate the dynamical behavior of the universe and the evolution of different cosmological parameters. The remainder of the paper is structured as follows. The fundamental formalism and the field equations of the $f(T,\mathcal{L}_m)$ gravity are determined in Sec.\ref{Basic formalisml}. The viscous fluid and the barotropic equation of state (EoS) are introduced in Sec.\ref{viscous with barotropic}. We constructed the Hubble parameter in the framework of $f(T,\mathcal{L}_m)$ gravity in Sec.\ref{f(T,L_m) Gravity}. We use the observational datasets to execute Bayesian MCMC analysis and constrain the free model parameters in Sec.\ref{observational analysis}. By assessing the behavior of the key cosmological quantities, we discuss the physical consequences of the proposed framework in  Sec.\ref{results}. In the end, Sec.\ref{conclusion} summarizes the conclusion of this investigation.
     
\section{Action Principles and Field Equations of $f(T,\mathcal{L}_m)$ gravity}
\label{Basic formalisml}

An alternate geometrical explanation of gravitation is offered by TG, in which the gravitational interaction is caused by torsion rather than curvature. The tetrad fields $(e^{A}_{\nu})$, which form an orthonormal basis in the tangent space at each point of spacetime, are the fundamental dynamical variables in this framework, and the spacetime metric can be defined from the tetrads as,

\begin{equation}
    g_{\mu\nu}=\eta_{ab}e^{a}_{\mu}e^{b}_{\nu},
\end{equation}

where $\eta_{ab}=\mathrm{diag}(1,-1,-1,-1)$ denotes the Minkowski metric. TG is formulated using the curvature-less Witzenb\"{o}ck connection \cite{aldrovandi2012teleparallel}, in contrast to GR, which uses the torsion-free Levi-Civita connection

\begin{equation}
    \Gamma^{\lambda}_{\mu\nu}
=e^{\lambda}_{a}\partial_{\nu}e^{a}_{\mu}.
\end{equation}

The corresponding torsion tensor is defined by

\begin{equation}
T^{\sigma}_{\mu\nu}
=\Gamma^{\sigma}_{\nu\mu}
-\Gamma^{\sigma}_{\mu\nu}
=e^{\sigma}{a}
\left(
\partial_{\mu}e^{a}_{\nu}
-\partial{\nu}e^{a}_{\mu}
\right).
\end{equation}

The contorsion tensor can be introduced from the torsion tensor as

\begin{equation}
    K^{\mu\nu}_{\ \ \sigma}
=-\frac{1}{2}
\left(
T^{\mu\nu}_{\ \ \sigma}
-T^{\nu\mu}_{\ \ \sigma}
-T{\sigma}^{\ \mu\nu}
\right),
\end{equation}

and the superpotential

\begin{equation}
    S_{\sigma}^{\ \mu\nu}
=\frac{1}{2}
\left(
K_{\sigma}^{\ \mu\nu}
+\delta_{\sigma}^{\mu}
T_{\alpha}^{\ \alpha\nu}
-\delta_{\sigma}^{\nu}
T_{\alpha}^{\ \alpha\mu}
\right).
\end{equation}

The Lagrangian density of TG is given by the torsion scalar

\begin{equation}
    T=S_{\sigma}^{\ \mu\nu}
T^{\sigma}_{\ \mu\nu},
\end{equation}

which can be explicitly expressed as,

\begin{equation}
    T=
\frac{1}{4}
T^{\sigma\mu\nu}T_{\sigma\mu\nu}
+\frac{1}{2}
T^{\sigma\mu\nu}T_{\nu\mu\sigma}
-T^{\sigma}_{\ \sigma\mu}
T^{\nu\mu}_{\ \ \nu}.
\end{equation}

The $f(T,\mathcal{L}_m)$ extension of TEGR is obtained by replacing the standard teleparallel Lagrangian with an arbitrary function involving the torsion scalar $T$ and the matter Lagrangian density $\mathcal{L}_m$. The action \cite{Cruz_2026,Bahamonde_2018} can be written as,

\begin{equation}\label{EH action}
    \frac{1}{2\kappa^2}
\int d^4xe
\left[
T+f(T,\mathcal{L}_m)
\right]
+\int d^4xe\mathcal{L}_m,
\end{equation}\
where $e=\det(e^{A}_{\ \nu})=\sqrt{-g}$ and
$\kappa^2=8\pi G = 1$. Varying the action(\ref{EH action}) with regards to the tetrad fields yields the gravitational field equations.

\begin{equation}
\begin{aligned}
&\left[e^{-1}\partial_\mu\!\left(ee_a^{\ \rho}S_\rho^{\ \mu\nu}\right)
-e_a^{\ \lambda}T^{\rho}_{\ \mu\lambda}S_\rho^{\ \nu\mu}\right](1+f_T) \\
&\quad
+e_a^{\ \rho}S_\rho^{\ \mu\nu}
\left(f_{TT}\partial_\mu T
+f_{T\mathcal{L}}\partial_\mu \mathcal{L}_m\right) \\
&\quad
-\frac{1}{4}f_{\mathcal{L}}
\left(e_a^{\ \rho}{T}{}_\rho^{\ \nu}
+e_a^{\ \nu}\mathcal{L}_m\right)
+e_a^{\ \nu}\left(\frac{T+f}{4}\right) \\
&\qquad
=4\pi G\,e_a^{\ \rho}{\mathcal{T}}{}_\rho^{\ \nu}.
\end{aligned}
\label{field equations}
\end{equation}

where

\begin{equation}
\begin{aligned}
f_T &= \frac{\partial f}{\partial T},
\qquad
f_{TT} = \frac{\partial^2 f}{\partial T^2},\\
f_{T\mathcal{L}_m}
&= \frac{\partial^2 f}{\partial T\,\partial \mathcal{L}_m},
\qquad
f_{\mathcal{L}_m}
= \frac{\partial f}{\partial \mathcal{L}_m}.
\end{aligned}
\end{equation}

We consider a spatially flat Friedmann-Lema$\Hat{i}$tre-Robertson-Walker spacetime in order to investigate cosmological dynamics.

\begin{equation}
ds^2=dt^2-a^2(t)\delta_{ij}dx^idx^j,
\end{equation}

with the diagonal tetrad $\mathrm{diag}(1,a,a,a)$ , where $a(t)$ or $a$ is the scale factor. For this geometry, the torsion scalar tales the simple form 

\begin{equation}
    T=-6H^2
\end{equation}

where $H=\dot a/a$ denotes the Hubble parameter. By evaluating the field equations with FLRW tetrad, the modified Friedmann equations can be obtained as

\begin{equation} \label{first friedmann equation}
H^2=
\frac{8\pi G}{3}\rho
-2H^2f_T
+\frac{1}{6}f_{\mathcal{L}_m}
(\rho+\mathcal{L}_m)
-\frac{f}{6},
\end{equation}

and

\begin{equation}
\begin{aligned}
\dot H
=&-4\pi G(\rho+p)
-\dot H\left(f_T-12H^2f_{TT}\right) \\
&-Hf_{T\mathcal{L}_m}\dot{\mathcal{L}}_m
-\frac{1}{2}f_{\mathcal{L}_m}(p+\mathcal{L}_m).
\end{aligned}
\label{second friedmann equation}
\end{equation}

The cosmological evolution in $f(T,\mathcal{L}_m)$ gravity is governed by these equations, which reduce to the standard teleparallel Friedmann equations in the limit $f(T,L_m)\rightarrow0$. Now, comparing the Eqs.\ref{first friedmann equation} and \ref{second friedmann equation} with standard GR

\begin{equation}
    3H^2 = 8 \pi G \left[ \rho + \rho_{DE} \right]
\end{equation}

and

\begin{equation}
    2 \dot{H} + 3 H^2 = -8 \pi G \left[p + p_{DE} \right]
\end{equation}

then DE sector energy density and pressure for $f(T,\mathcal{L}_m)$ gravity can be obtained as

\begin{equation}
    \rho_{DE} = \frac{3}{8 \pi G} \left[ -2 H^2 f_T + \frac{1}{6} f_{\mathcal{L}} \left( \rho + \mathcal{L}_m \right) - \frac{f}{6} \right] 
\end{equation}

and

\begin{equation}
\begin{aligned}
p_{DE}
=&-\rho_{DE}
+\frac{1}{8\pi G}
\Big[
2\dot{H}\left(f_T-12H^2f_{TT}\right) \\
&\qquad
+2H\,\partial_T\!\left(\mathcal{L}_m\right)f_{T\mathcal{L}}
+f_{\mathcal{L}}\left(p+\mathcal{L}_m\right)
\Big].
\end{aligned}
\end{equation}

\section{Bulk Viscous Cosmology with a Barotropic Equation of State}\label{viscous with barotropic}

This section explains how bulk viscosity is incorporated into the cosmological fluid in the framework of $f(T,\mathcal{L}_m)$ gravity, and how it modifies the Friedmann equations and the effective energy density and pressure.\\
In order to study the cosmic dynamics in $f(T,\mathcal{L}_m)$ gravity, it is assumed that the universe is filled with a homogeneous and isotropic viscous fluid. Since dissipative processes naturally take place with the expansion of the universe, the assumption of a perfect fluid in cosmology is an idealization. These irreversible thermodynamic processes can be significantly demonstrated through viscosity, which has been thoroughly investigated in both GR and modified theories of gravity \cite{Eckart1940,Israel1976,IsraelStewart1979,Brevik2005,Brevik2006}. The energy-momentum tensor of a viscous fluid is be given as

\begin{equation}
    \mathcal{T}_{\mu\nu}=(\rho+p_{\rm viscosity})u_\mu u_\nu-p_{\rm eff} g_{\mu\nu},
\end{equation}

Here, $\rho$ is ordinary cosmic-fluid energy density, $u^\mu$ is the four-velocity satisfying $u^\mu u_\mu=1$, $p_{viscosity}$ represents pressure of the fluid that possesses bulk viscosity $p_{\rm eff}$ denotes the effective pressure. In an FLRW universe, shear viscosity vanishes due to spatial theory, leaving viscosity as the only relevant dissipative contribution. As a result, the bulk-viscosity pressure is modified as

\begin{equation}\label{viscous fluid}
    p_{\rm viscosity}=p-3\xi H
\end{equation}

where $\xi$ denotes the viscosity coefficient, while $H=\dot a/a$ is the Hubble expansion rate. The observed late-time accelerated cosmic expansion may be attributed to the viscous term, which introduces an effective negative pressure that can influence the expansion dynamics.\\

In the current study, we consider the simplest physically motivated choice where the viscosity coefficient stays constant over the course of cosmic evolution

\begin{equation}
    \xi=\xi_0,\qquad \xi_0>0.
\end{equation}

Consequently, the effective pressure is modified to

\begin{equation} \label{viscous pressure with constant viscosity}
	p_{\rm viscosity}=p-3\xi_0H.
\end{equation} 

We assume that the cosmic fluid follows the barotropic equation of state through an arbitrary pressure, 

\begin{equation}
	p=\omega \rho
\end{equation}

where $\omega$ is the barotropic equation-of-state parameter. During cosmic expansion, the viscosity term $3 \xi_0 H$ acts as an additional negative pressure produced by irreversible thermodynamic process. Without adding an explicit DE component, this dissipative contribution offers an alternative mechanism for driving the late-time accelerated expansion and has the potential to drastically modify the cosmological dynamics. From the $\omega$CDM cosmology, we have

\begin{equation}\label{wcdm 1}
    3 H^2 = \rho_m + \rho_{eff(DE)}
\end{equation}

and 
\begin{equation} \label{wcdm 2}
    \dot{H} = -\frac{1}{2} \left[\rho_m + \rho_{eff(DE)} + p_{eff(DE)} \right]
\end{equation}

Now comparing Eqs,\ref{first friedmann equation}, \ref{second friedmann equation}, \ref{wcdm 1}, and \ref{wcdm 2} the effective DE density and effective DE pressure can be obtained as

\begin{equation}
    \rho_{eff(DE)} = \rho + T f_T + \frac{1}{2} \left[f_\mathbf{L} \left(\rho + \mathcal{L}_m \right)-f \right] - \rho_m
 \end{equation}

 and

 \begin{equation}
\begin{aligned}
p_{eff(DE)}
=&\left[
\frac{(\rho+p_{eff})
+2Hf_{T\mathcal{L}}\dot{\mathcal{L}}_m
+f_{\mathcal{L}}(p_{eff}+\mathcal{L}_m)}
{1+f_T+2Tf_{TT}}
\right] \\
&-\left[
\rho+Tf_T
+\frac{1}{2}
\left(
f_{\mathcal{L}}(\rho+\mathcal{L}_m)-f
\right)
\right].
\end{aligned}
\end{equation}
where, $\rho_{m}$ represents the ordinary matter density.

\section{Reconstructing the Cosmic Expansion History in $f(T,\mathcal{L}_m)$ Gravity} \label{f(T,L_m) Gravity}

In this framework, to investigate the cosmological evolution we consider $f(T,\mathcal{L}_m) = \alpha T + \beta \mathcal{L}_m$, where $\mathcal{L}_m=\rho$. From this mentioned form, GR can be recovered for $\alpha \rightarrow 0$ and $\beta \rightarrow 0$. Moreover, using this form in the Friedmann Eq.(\ref{first friedmann equation}), the effective energy density will become

\begin{equation}\label{density calculation}
    \rho_{eff} = 6 H^2 \left(\frac{1 + \alpha}{2 + \beta} \right)
\end{equation}

At the present time, it becomes $\rho_0 = 6 H_0^2 \left(\frac{1 + \alpha}{2 + \beta} \right)$ and comparing this with Eq.(\ref{density calculation}), the final form of the effective energy density can be obtained as

\begin{equation}\label{final form of density}
    \rho_{eff} = \rho_0 \left(\frac{H}{H_0}\right)^2
\end{equation}

where $H_0$ is the present value of the Hubble parameter. Taking Friedmann Eq.(\ref{second friedmann equation}) into account, we obtain the following expression for the time derivative of the Hubble parameter

\begin{equation}
    \dot{H} \left(\frac{1 + \alpha}{1 + \beta}\right) = - \frac{1}{2} \left( \rho + p_{eff} \right)
\end{equation}

Considering Eq.(\ref{final form of density}) and the viscous fluid (\ref{viscous pressure with constant viscosity}), we get the expression

\begin{equation} \label{diff. eq. of H}
    \frac{\dot{H}}{H^2} - \frac{3}{2} \xi_0 \left( \frac{1 + \beta}{1 + \alpha} \right) \frac{H}{H^2} = - \frac{1}{2} \left( \frac{1 + \beta}{1 + \alpha} \right) \frac{\rho_0}{H_0^2} \left(1 + \omega \right)
\end{equation}

This is a first-order differential equation, and to solve this, we convert it into $\frac{1}{H} \frac{dH}{dt} = \frac{dH}{d \left(lna \right)}$ with $lna$ as the e-folding time and $a$ as the scale factor. Now. solving the Eq.(\ref{diff. eq. of H}) for $a = \left(1+z \right)^{-1}$ the final explicit expression of the Hubble model is

\begin{equation}
\begin{aligned}
H(z)
=&\,H_0(1+z)^{\frac{3(1+\beta)(1+\omega)}{2+\beta}} \\
&+\frac{\xi_0(2+\beta)}
{2(1+\alpha)(1+\omega)}
\left[
1-(1+z)^{\frac{3(1+\beta)(1+\omega)}{2+\beta}}
\right].
\end{aligned}
\label{Hubble model}
\end{equation}

Thus, the expansion history in the framework of $f(T,\mathcal{L}_m)$ gravity with a barotropic viscous fluid is provided by the obtained Hubble parameter (\ref{Hubble model}). By construction, the solution obeys a normalization condition $H(z=0) = H_0$, where $H_0$ denotes the present-day values of the Hubble parameter. The barotropic EoS parameter is denoted by $\omega$. The gravitational and torsion-matter coupling sectors of the theory are characterized by the parameters $\alpha$ and $\beta$, respectively. The constant viscosity coefficient $\xi_0$ generates a negative effective  pressure that can modify the cosmic expansion history at low redshifts.\\

Various significant limiting situations are naturally recovered by this derived solution. The Hubble parameter is getting reduced to non-viscous $f(T,\mathcal{L}_m)$ cosmology, when the viscosity coefficient $\xi_0$ is absent. The torsion-matter coupling is eliminated by considering $\beta=0$, yielding $H(z)=H_0(1+z)^{\frac{3}{2}(1+\omega)}$, which is precisely the standard FLRW solution for GR with barotropic fluid and also, in particular, this expression is reduced to Einstein's de-sitter matter-dominated universe for $\omega = 0$, $H(z)=H_0(1+z)^{3/2}$.\\

In this study, all the model parameters $H_0$, $\xi_0$, $\omega$, $\alpha$, and $\beta$ are treated as the free parameters and constrained using the Bayesian statistical inference by using the MCMC technique based on available observational datasets. Therefore, the suggested model can closely reproduce the expansion history over the accessible redshift range for suitable values of the model parameters.

\section{Bayesian Observational Constraints and Parameter Estimation}\label{observational analysis}

To investigate whether the proposed viscous $f(T,\mathcal{L}_m)$ model is consistent with recent observations, its parameters are constrained by using the recent cosmological observations, such as the observational Hubble ($H(z)$) measurements, the Pantheon+SH0ES Type Ia supernova compilation, DESI DR II baryon acoustic oscillation (BAO) data, and the cosmic wave background (CMB). The complementary datasets offer stringent constraints on the model parameters and probe the universe's expansion history across a wide redshift range. The MCMC method is used to estimate free model parameters within a Bayesian framework. The Hubble parameter is then reconstructed using the resulting best-fit values, enabling a direct comparison between the theoretical predictions and the observational data.    

\subsection*{Observational Datasets} \label{datasets}

\subsubsection{\textbf{Observational $H(z)$ Data}}
The observational $H(z)$ measurements are a crucial datasets for constraining cosmological models and offer a direct explanation of the expansion history of the cosmos. The collection of 32 cosmic chronometer measurements reported in \cite{BHAGAT2023101250} is utilized in this investigation. To carry out the estimation process within the Bayesian framework using the MCMC technique, the $\chi ^2$ statistic for this datasets can be expressed as

\begin{equation}
    \chi^{2}_{\mathrm{\textit{H(z)}}} = \sum_{i=1}^{32} \frac{\left[ H_{\mathrm{th}}(z_i) - H_{\mathrm{obs}}(z_i) \right]^2}{\sigma_i^2}\,,
\end{equation}

where $H_{\rm th}(z_i)$ and $H_{\rm obs}(z_i)$ express the theoretical and observed values of the Hubble parameter at $z_i$, respectively, whereas $\sigma_i$ is the corresponding observational uncertainty linked with each measurement.  

\subsubsection{\textbf{Pantheon+SH0ES Type Ia Supernova Data}}
One of the most efficient cosmic probes for exploring the late-time expansion of the universe is the Type Ia supernova (SNe Ia). They serve as standard candles because of their nearly uniform intrinsic luminosity, which allow precise luminosity distance measurements across a wide redshift range. In this study, we use the Pantheon+SH0ES compilation, which includes 1701 light-curve measurements from 1550 spectroscopically confirmed SNe Ia covering the redshift range $0.00122<z<2.2613$ \cite{Brout_2022}.\\

The theoretical distance modulus for a certain cosmological model can be expressed as

\begin{equation}
\mu_{\mathrm{th}}(z,\theta) = 5 \log_{10}\left(d_L(z,\theta)\right) + 25,
\end{equation}

where $\boldsymbol{\theta}$ denotes the free model parameters vector and the luminosity distance is determined by

\begin{equation}
d_L(z,\theta) = (1+z)c \int_{0}^{z} \frac{dz'}{H(z',\theta)},
\end{equation}

Here, $c$ is the speed of light in vacuum and $H(z,\boldsymbol{\theta})$ is the theoretical Hubble parameter predicted by the model. The Pantheon+SH0ES datasets offers the observed distance modulus and the full covariance matrix with statistical and systematic uncertainties. Accordingly, to carry out the estimation process, the $\chi ^2$ function can be constructed as

\begin{equation}
    \chi^2_{\mathrm{SN}} = \Delta \mu^{T} C^{-1} \Delta \mu,
\end{equation}

where the residual vector is where, $\Delta \mu_j = \mu_{\mathrm{th}}(z_j,\theta) - \mu_{\mathrm{obs}}(z_j)$. The superscript $T$ denotes the transpose of the residual vector, and $\mathbf{C}$ represents the covariance matrix of the Pantheon+SH0ES model.

\subsubsection{\textbf{DESI DR II Baryon Acoustic Oscillation Dataset}}
One of the most effective geometrical probes for analyzing the expansion history of the universe is the Baryon Acoustic Oscillations (BAO). The co-moving sound horizon evaluated at the baryon drag epoch, or $r_d$, is a characteristic scale and is used as a standard ruler for constraining cosmological models, expressed by

\begin{equation}
r_d = \int_{z_d}^{\infty} \frac{c_s(z)}{H(z)} \, dz,
\end{equation}

where $H(z)$ is the Hubble parameter, $c_s (z)$ is the sound speed of the photon-baryon fluid and $z_d$ denotes the redshift of baryon drag epoch. We employ the latest DESI DR II BAO observations \cite{AbdulKarim_2025} in this study, which offer precise constraints on the expansion history over the redshift range $0.295<z<2.33$. The dataset report dimensionless distance indicators $D_M/r_d$, $D_V/r_d$, and $D_H/r_d$, defined as

\begin{equation}
\begin{aligned}
\frac{D_M}{r_d}
&=(1+z)^{-1}\frac{D_L}{r_d},
\qquad
\frac{D_V}{r_d}
=\frac{\left(cD_L\frac{z}{H(z)}\right)^{1/3}}{r_d},\\
\frac{D_H}{r_d}
&=\frac{c}{r_dH(z)}.
\end{aligned}
\end{equation}

where $D_M$, $D_V$, and $D_H$ refer to Hubble distance, the volume-averaged distance, and the transverse co-moving distance, respectively. The agreement between the DESI DR II BAO observations and theoretical predictions is quantified through the $\chi ^2$ statistic

\begin{equation}
    \chi^2_{DESI}
=
\Delta^{\mathrm T}
C_{DESI}^{-1}
\Delta,
\end{equation}
 where $C_{\rm DESI}$ is the covariance matrix of the DESI DR II BAO dataset, which contains both statistical and systematic uncertainties, and $\Delta$ is the residual vector produced by the difference between the theoretical predictions and the corresponding observational measurements.

 \subsubsection{\textbf{Cosmic Microwave Background Dataset}}

 The $f(T,\mathcal{L}_m)$ cosmological model has a largely geometric impact on the cosmic microwave background (CMB) because it mostly modifies the late-time expansion history of the universe. Therefore, we use the compressed CMB likelihood, which efficiently captures the geometric information relevant for background cosmology while minimizing the impact of small-scale effects and potential systematic uncertainties, instead of using the full CMB temperature and polarization power spectra. In this research, we employ the commonly used CMB distance priors from the Planck measurements \cite{Aubourg2015}. The compressed likelihood is expressed by the parameter set ${R,l_A,\omega_b}$ \cite{Escamilla2024} and defined as

\begin{equation}
\begin{aligned}
R
&=\sqrt{\Omega_{m0}}\,H_0D_M(z_*),\\
l_A
&=\pi\frac{D_M(z_*)}{r_s(z_*)},\\
\omega_b
&=\Omega_bh^2.
\end{aligned}
\end{equation}

where $D_M(z_*)$ refers to the co-moving distance of the surface last-scattering, $r_s(z_*)$ is the co-moving acoustic horizon evaluated at the decoupling epoch, and $Z_*$ is the photo decoupling redshift. The distance priors efficiently encode the geometrical information of the CMB and offer a computationally effective way to incorporate early-Universe constraints into the parameter estimation. The corresponding CMB $\chi ^2$ statistic is given by

\begin{equation}
    \Delta^{T}
C_{\rm CMB}^{-1}
\Delta,
\end{equation}

where $C_{\rm CMB}$ is the corresponding covariance matrix, and $\Delta$ is the residual vector between the observed CMB distance priors and the theoretical predictions. \\

To obtain the allowed parameter space of the proposed cosmological Hubble model, we execute different combinations of the observational datasets in a joint Bayesian analysis. Three independent dataset combinations are taken into account in this study: (i) $H(z) $ + Pantheon+SH0ES, (ii) DESI DR2 BAO + CMB, and (iii) $H(z)$ + Pantheon+SH0ES + DESI DR II BAO + CMB. Consequently, the corresponding $\chi ^2$ statistics are expressed as

\begin{equation}
\begin{aligned}
\chi^2_{H(z)+\rm SN}
&=\chi^2_{H(z)}
+\chi^2_{\rm SN},\\[2mm]
\chi^2_{\rm DESI+\rm CMB}
&=\chi^2_{\rm DESI}
+\chi^2_{\rm CMB},\\[2mm]
\chi^2_{\rm Combined}
&=\chi^2_{H(z)}
+\chi^2_{\rm SN}
+\chi^2_{\rm DESI}
+\chi^2_{\rm CMB}.
\end{aligned}
\end{equation}

The likelihood function corresponding to each dataset combination is related to the total chi-square through

\begin{equation}
   \ln{\mathcal{L}} = -\frac{1}{2}\chi^2,
\end{equation}

where $\chi ^2$ denotes the respective chi-square combination. The Bayesian MCMC technique is used to maximize the likelihood fr constraining the free parameters. This approach yields the best-fit values, posterior distributions, and associated confidence intervals. To explore the posterior distribution of the parameter space, we consider the uniform priors, and the details are given in the Table-\ref{tab:priors}.

\begin{table}[htbp]
\centering
\renewcommand{\arraystretch}{1.3}
\setlength{\tabcolsep}{12pt}
\captionsetup{justification=raggedright,singlelinecheck=false}
\caption{Uniform prior ranges considered for the free model parameters.}
\label{tab:priors}

\begin{tabular}{lc}
\toprule
\textbf{Parameter} & \textbf{Prior Range} \\
\hline
$H_0$      & $(65,\,80)$ \\
$\alpha$   & $(-5,\,-1)$ \\
$\beta$   & $(-5,\,-2)$ \\
$\omega$    & $(-1.3,\,0.3)$ \\
$\xi_0$    & $(0,\,100)$ \\
\hline
\end{tabular}
\end{table}

\begin{figure*}[!t]
\centering
\includegraphics[width=0.65\textwidth]{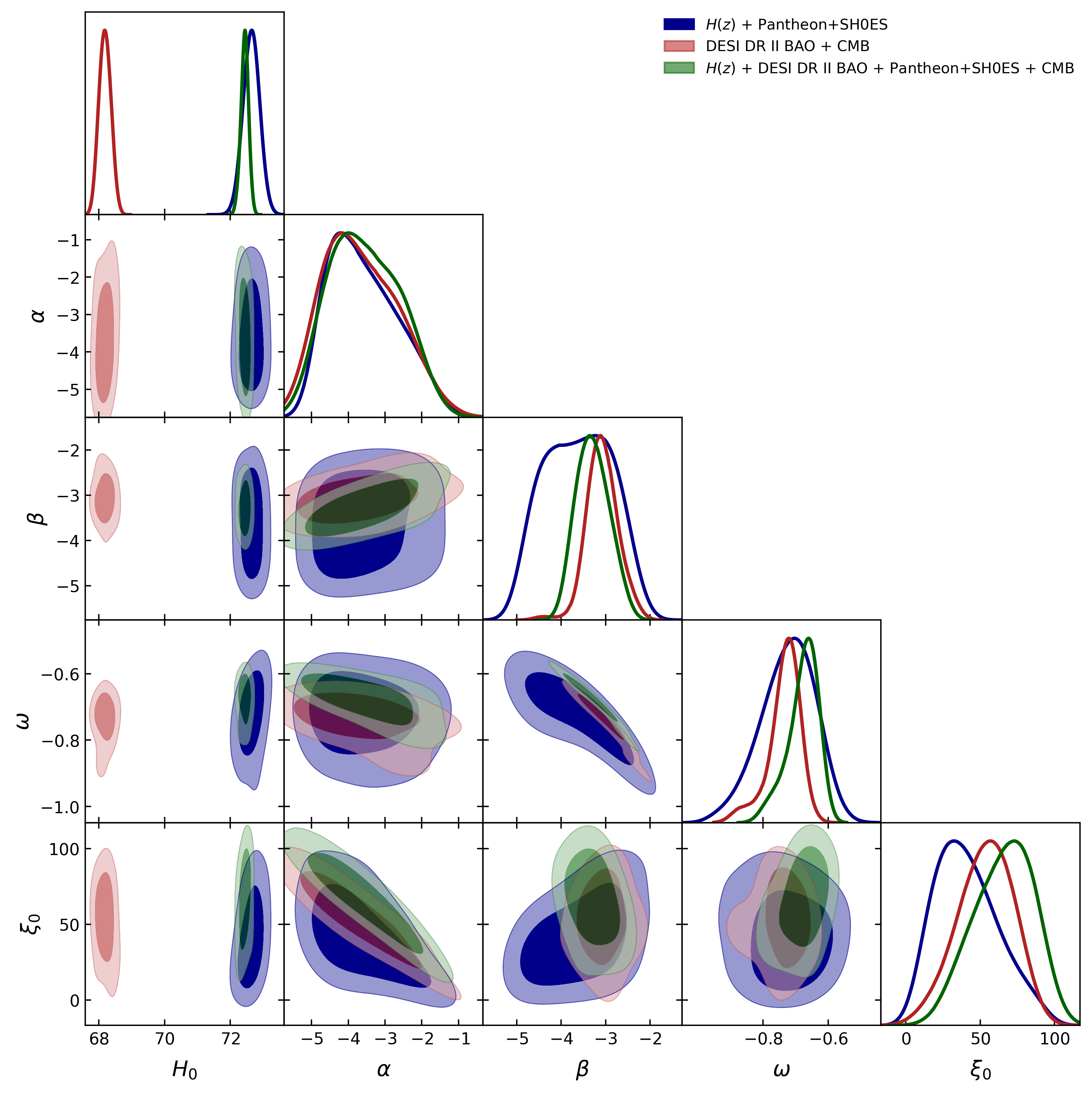}
\caption{\raggedright Two-dimensional confidence contour plots for the model parameters $H_0$, $\alpha$, $\beta$, $\omega$, and $\xi_0$ derived from the combined $H(z)$ + Pantheon+SH0ES, DESI DR II BAO + CMB, and $H(z)$ + Pantheon+SH0ES + DESI DR II BAO + CMB (Combined) analyses. The shaded contours represent the $1\sigma$ and $2\sigma$ confidence regions.}
\label{fig:triangle_plot}
\end{figure*}

Fig.\ref{fig:triangle_plot} the two dimensional marginalized posterior distributions of the free model parameters $(H_0,\alpha,\beta,\omega,\xi_0)$ obtained from the Bayesian MCMC analysis using $H(z)$ + Pantheon+Sh0ES, DESI DR II BAO + CMB, and the combined $H(z)$ + Pantheon+SH0ES + DESI DR II BAO + CMB datsets. The one-dimensional marginal posterior distributions of individual parameters are presented in the diagonal panels, and the associated two-dimensional confidence contours are shown in the off-diagonal panels. The $68 \%$ ($1\sigma$) and $95 \%$ ($2\sigma$) confidence levels are shown by the inner and outer contours, respectively. While the confidence contours illustrate the correlations between the free parameters and indicate the statistically acceptable regions of the parameter space, the posterior distributions present well-constrained estimates of the model parameters. Table-\ref{tab:confidence_intervals} illustrates the associated best-fit values together with their $1\sigma$, $2\sigma$, and $3\sigma$ confidence intervals.

\begin{table*}[htbp]
\centering
\renewcommand{\arraystretch}{1.5}
\setlength{\tabcolsep}{3pt}
\captionsetup{justification=raggedright,singlelinecheck=false}
\caption{Best-fit values with $1\sigma$ and $2\sigma$ confidence intervals for the model parameters $H_0$, $\omega$, $\alpha$, $\beta$, and $\xi_0$ obtained from the $H(z)$ + Pantheon+SH0ES, DESI DR II BAO + CMB, and the combined $H(z)$ + Pantheon+SH0ES + DESI DR II BAO + CMB datasets.}
\label{tab:confidence_intervals}

\begin{tabular}{llccccc}
\toprule
\textbf{Dataset} & & \boldmath{$H_0$} & \boldmath{$\omega$} & \boldmath{$\alpha$} & \boldmath{$\beta$} & \boldmath{$\xi_0$} \\
\hline

{$H(z)$ + Pantheon+SH0ES}
& $1\sigma$
& $72.64730^{+0.25044}_{-0.25836}$
& $-0.71689^{+0.07922}_{-0.09307}$
& $-3.73978^{+1.21148}_{-0.89581}$
& $-3.60442^{+0.85291}_{-0.90895}$
& $38.43919^{+25.67625}_{-20.32827}$ \\

& $2\sigma$
& $72.64730^{+0.49950}_{-0.51373}$
& $-0.71689^{+0.14533}_{-0.19536}$
& $-3.73978^{+2.17760}_{-1.20578}$
& $-3.60442^{+1.35338}_{-1.31827}$
& $38.43919^{+51.57274}_{-32.87652}$ \\

\hline

{DESI DR II BAO + CMB}
& $1\sigma$
& $68.18475^{+0.19394}_{-0.18357}$
& $-0.72599^{+0.03169}_{-0.04157}$
& $-3.85301^{+1.29873}_{-0.89822}$
& $-3.08779^{+0.30343}_{-0.29738}$
& $55.76580^{+17.96177}_{-20.79864}$ \\

& $2\sigma$
& $68.18475^{+0.36715}_{-0.34502}$
& $-0.72599^{+0.06333}_{-0.15897}$
& $-3.85301^{+2.38572}_{-1.11425}$
& $-3.08779^{+0.80913}_{-0.66454}$
& $55.76580^{+31.82642}_{-45.75405}$ \\

\hline

{Combined}
& $1\sigma$
& $72.45158^{+0.10100}_{-0.12174}$
& $-0.66640^{+0.03779}_{-0.05953}$
& $-3.69484^{+1.24499}_{-0.95442}$
& $-3.31311^{+0.43921}_{-0.40486}$
& $68.97655^{+18.17734}_{-25.66935}$ \\

& $2\sigma$
& $72.45158^{+0.23066}_{-0.26709}$
& $-0.66640^{+0.05470}_{-0.13126}$
& $-3.69484^{+1.85115}_{-1.25263}$
& $-3.31311^{+0.77941}_{-0.60346}$
& $68.97655^{+28.76839}_{-44.13441}$ \\

\hline
\end{tabular}
\end{table*}

From the table, it can be observed that, depending on the dataset combination, the constrained Hubble constant ($H_0$) values vary range $(68.18,72.65)$, demonstrating agreement with current observational predictions of the current expansion rate. Additionally, the barotropic EoS parameter is consistently constrained to $\omega\approx-0.7$, indicating that the cosmic fluid possesses a negative pressure. Such a value is consistent with an effective fluid that may govern the cosmological dynamics, and it lies inside the quintessence regime $(-1<\omega<-1/3)$. Results of the observational analyses further support the internal consistency of the proposed cosmological model by imposing well-defined constraints on the remaining model parameters. 

\begin{figure}[htbp] 
   \centering 
   \mbox{\includegraphics[scale=0.4]{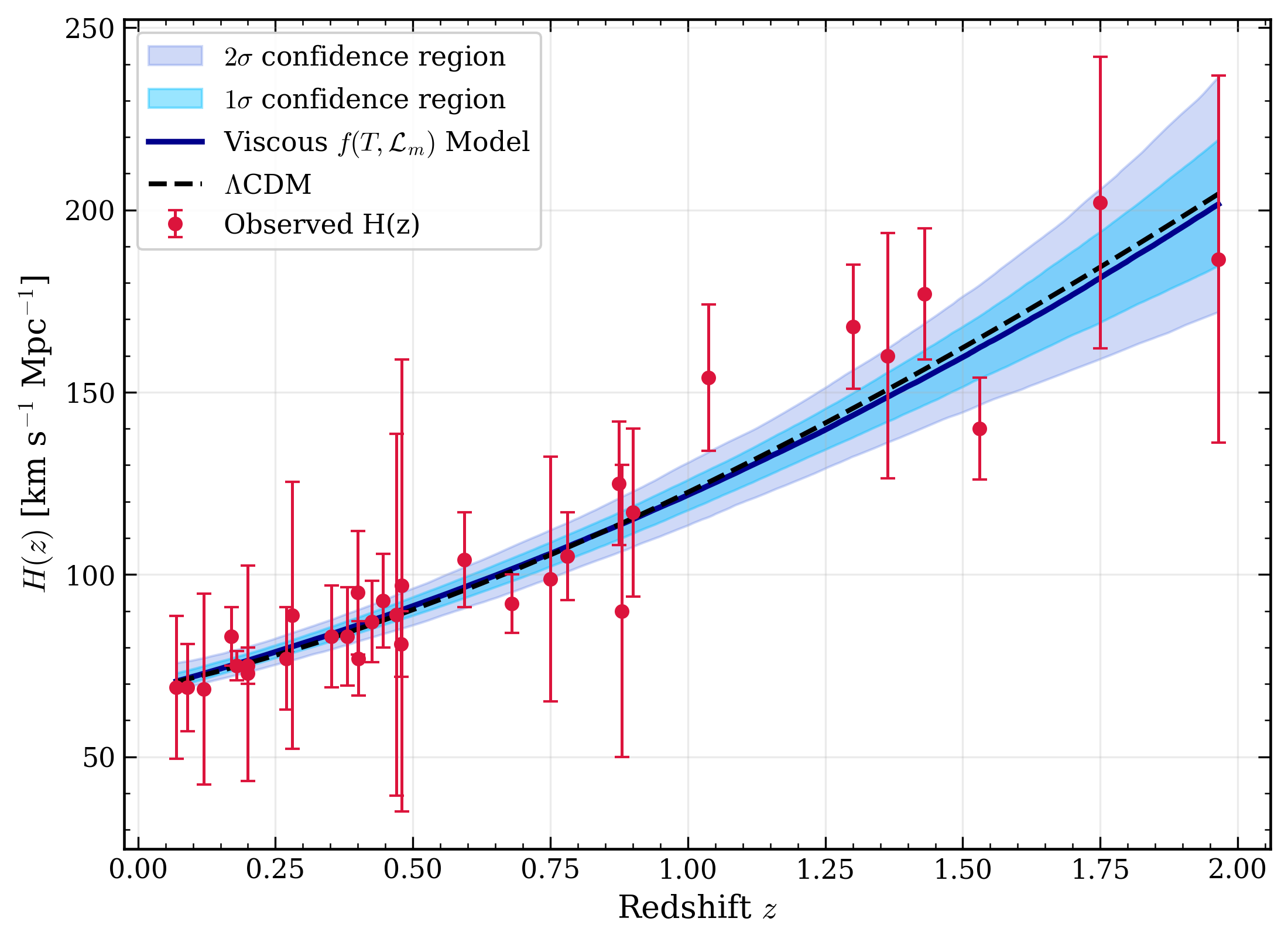}}   
    \hspace{10px}
    \mbox{\includegraphics[scale=0.4]{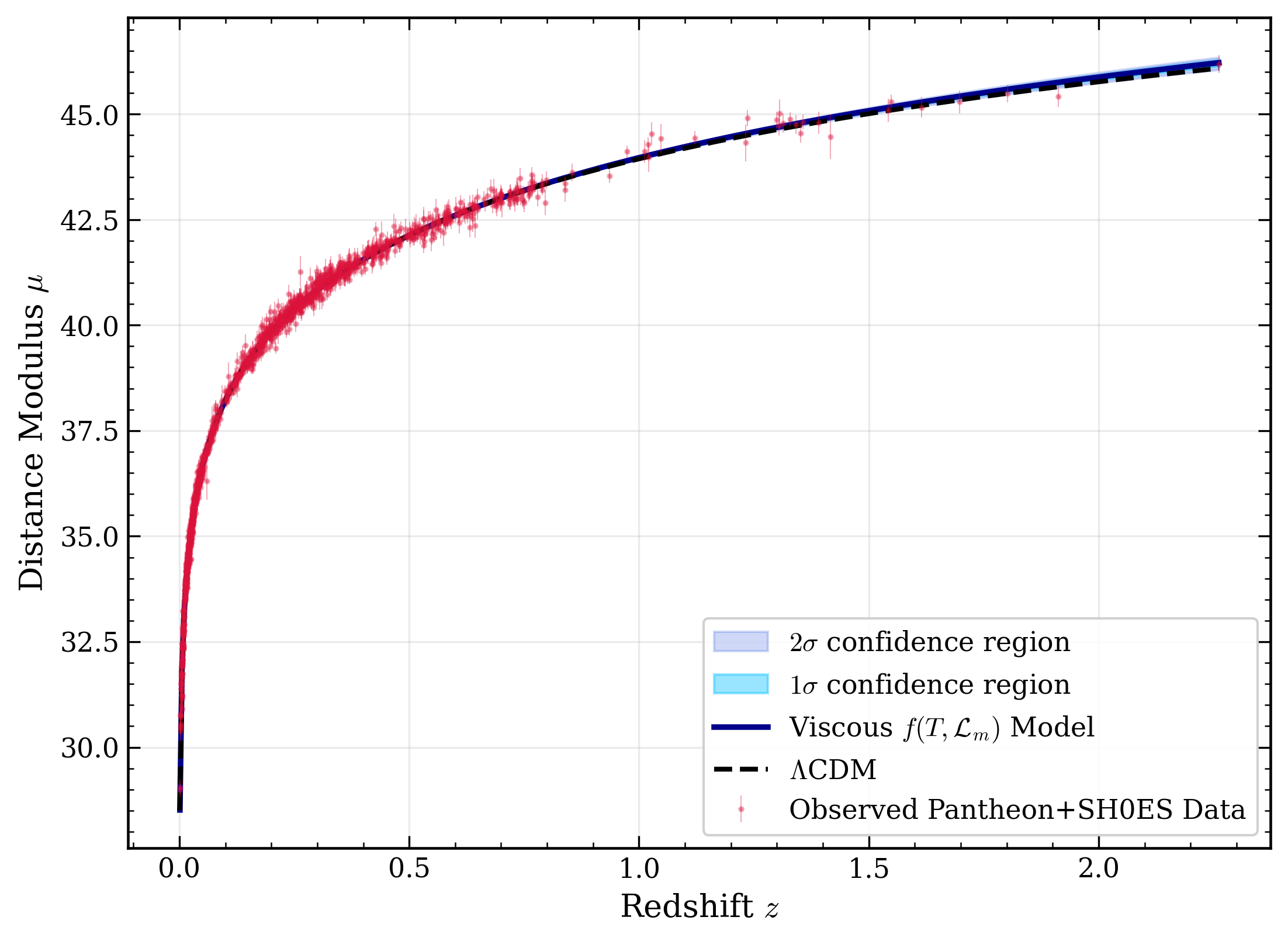}}
    \caption{\raggedright (i) Error-bar representation of the Hubble parameter ($H(z)$) (upper panel). (ii) Reconstructed distance modulus ($\mu$) versus redshift (lower panel).} 
   \label{Curve Fitting}
\end{figure}

Fig.\ref{Curve Fitting} illustrates the reconstructed Hubble parameter and distance modulus corresponding to the best-fit parameter values from $H(z)$ and Pantheon+SH0ES datasets. In both scenarios, the theoretical predictions of the proposed viscous $f(T,\mathcal{L}_m)$ cosmological model are displayed together with the corresponding observational data and the standard $\Lambda$CDM model. The shaded regions correspond to the $1 \sigma$ and $2 \sigma$ Bayesian confidence regions derived through MCMC sampling. The reconstructed curves are found in good agreement with the observational data over the entire redshift range and closely mimic the $\Lambda$CDM evolution, indicating that the proposed model successfully describes the background cosmological observations.

\section{Cosmological Dynamics and Observational Implications}\label{results}
 
In this section, we use the constrained values of the free parameters from the Bayesian MCMC analysis to investigate the evolution of cosmological parameters (both geometrical and physical) in order to understand the dynamical behaviour of the universe within this framework and their agreement with observational data. These quantities are analyzed through their respective redshift evolution and offer significant insights into the role of the viscosity and maximal torsion-matter coupling in shaping the cosmic evolution.  

\begin{figure}[htbp] 
   \centering 
   \mbox{\includegraphics[scale=0.65]{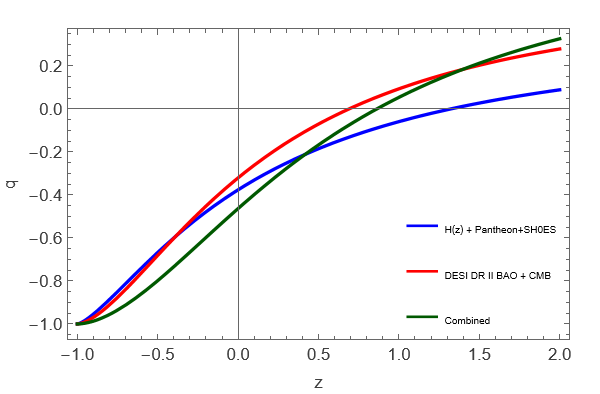}}   
    \caption{\raggedright Plot of the Deceleration Parameter for $H(z)$ + Pantheon+SH0ES. DESI DR II BAO + CMB, and combined datasets.} 
   \label{deceleration parametr}
\end{figure}

\begin{figure}[htbp] 
   \centering 
    \mbox{\includegraphics[scale=0.65]{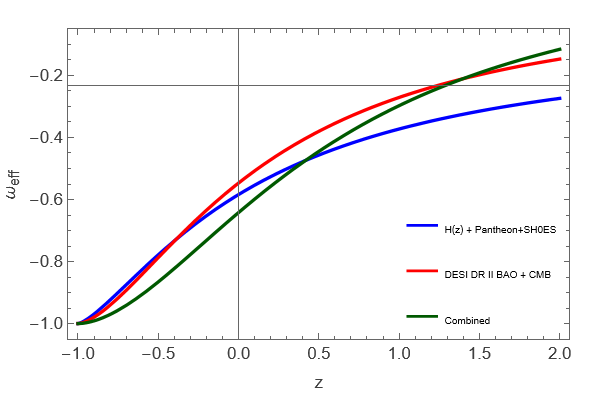}}
    \hspace{10px}
    \mbox{\includegraphics[scale=0.65]{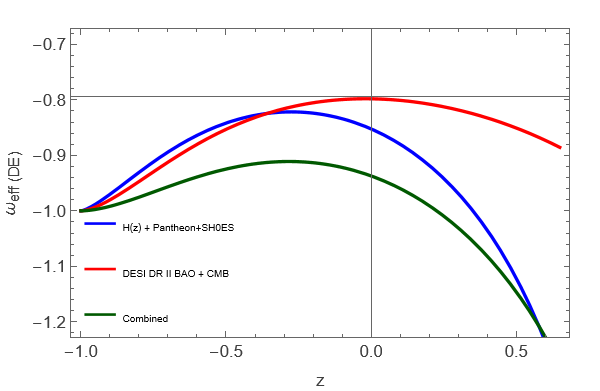}}
    \caption{\raggedright (i) Effective equation-of-state parameter (upper panel). (ii) Effective dark energy equation-of-state parameter (lower panel).} 
   \label{eos parametrs}
\end{figure}

\begin{figure}[htbp] 
   \centering 
   \mbox{\includegraphics[scale=0.65]{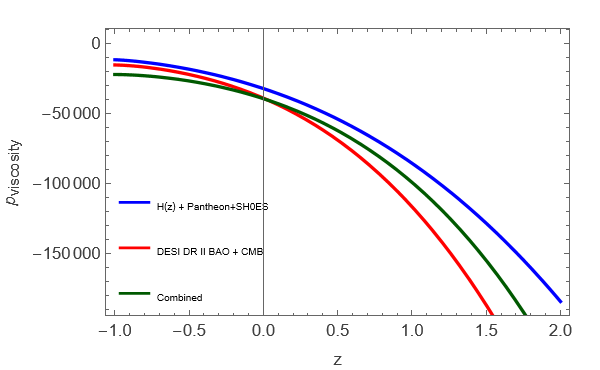}}   
    \hspace{10px}
    \mbox{\includegraphics[scale=0.65]{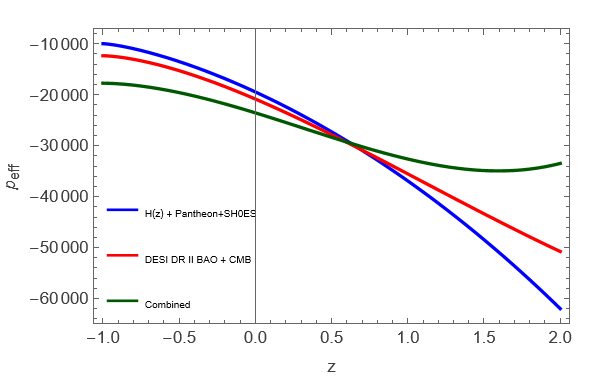}}
    \caption{\raggedright Evolution of the viscous pressure (upper panel). (ii) Evolution of the effective pressure (lower panel).} 
   \label{pressures}
\end{figure}

From the Fig.\ref{deceleration parametr}, the deceleration parameter clearly depicts the transitional behavior from early deceleration to the present acceleration. For $H(Z)$ Pantheon+SH0ES, DESI DR II BAO + CMB, and combined datasets, the transitions occur at the redshifts $z_t = 1.34$, $z_t = 0.69$, and $z_t = 0.86$, respectively. The universe is now going through accelerated expansion, evidenced by the corresponding current values of the deceleration parameter, which are around $q_0  = -0.38$, $q_0 = -0.32$, and $q_0 = -0.46$ and in agreement with the literature \cite{MYRZAKULOV2025116916, Wang2023}. The effective EoS parameter ($\omega_{eff}$) and effective DE EoS ($\omega_{de(eff)}$) are displayed in the Fig.\ref{eos parametrs}, where the former is shown in the upper panel and the latter appears in the lower panel. It is observed that $\omega_{eff}$ approaches the quintessence regime at late times and stays completely negative throughout the considered redshift range and giving present values around $-0.586$, $-0.548$, and $-0.643$ for $H(z)$ + Pantheon+SH0ES, DESI DR II BAO + CMB, and combined datasets, respectively. In a comparable manner, $\omega_{de(eff)}$ depicts the quintessence-like behavior late universe and remains negative throughout cosmic history, having present values around $-0.853$, $-0.798$, and $-0.937$ for $H(z)$ + Pantheon+SH0ES, DESI DR II BAO + CMB, and combined datasets, respectively. \\

The evolution of the viscous pressure ($p_{visco}$) and the effective pressure ($p_{eff}$) for the $H(z)$ + Pantheon+SH0ES, DESI DR II BAO + CMB , and combined datasets are displayed in the Fig.\ref{pressures}. For all dataset combinations, it is observed that both quantities stay completely negative across the considered redshift interval. While the negative viscous pressure demonstrates the dissipative contribution caused by viscosity, the constantly negative effective pressure indicates that the total cosmic fluid exhibits a negative-pressure behavior during its evolution. Furthermore both pressures depict smooth and abrupt transition-free evolution, suggesting the dynamical stability and physical viability of the suggested viscous $f(T,\mathcal{L}_m)$ cosmological model throughout the whole redshift range under consideration.\\

\begin{figure*}[!t]
\centering
\includegraphics[width=0.65\textwidth]{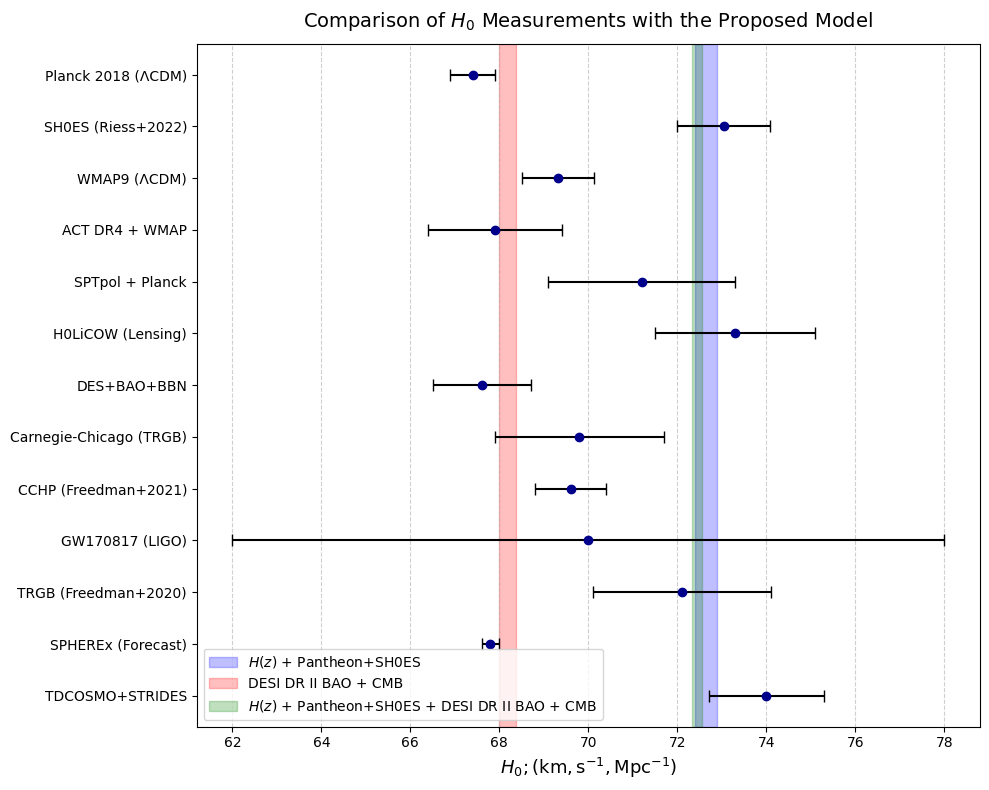}
\caption{\raggedright Comparison of the best-fit values of the Hubble constant, $H_0$, obtained from the proposed viscous $f(T,\mathcal{L}_m)$ model with several independent observational measurements corresponding to the $H(z)$ + Pantheon+SH0ES, DESI DR II BAO + CMB, and combined datasets.}
\label{fig:whisker_plot}
\end{figure*}

The Hubble constant value obtained from the current analysis is compared with various independent cosmological observations in the Fig.\ref{fig:whisker_plot}. The vertical shaded bands indicate the $1 \sigma$ confidence intervals corresponding to the three observational datasets combinations studied in this work namely, $H(z)$ + Pantheon+SH0ES, DESI DR II BAO + CMB, and combined analysis. The estimated values of $H_0$ are found to be within the range $(68.18–72.65)$, depending on the dataset combinations, highlighting the sensitivity of the inferred Hubble constant to various observational probes. The comparison also demonstrates that the derived constraints remain compatible with multiple early-universe determinations and are consistent with various independent late- and intermediate-time measurements within their specified uncertainties. Overall, the suggested viscous $f(T,\mathcal{L}_m)$ cosmological model yields observationally viable estimates of the current Hubble constant across various cosmological datasets.
 
\section{Conclusion}\label{conclusion}

In this study, we have used a barotropic EoS for the cosmic fluid to examine the cosmological dynamics of viscous fluid in $f(T,\mathcal{L}_m)$ gravity framework. Together with the viscous contribution,  the maximal coupling between torsion and matter Lagrangian provides a broader description of the cosmic fluid that can modify the late-time expansion of the universe history. We employed three different combinations of observational datasets $H(z)$ + Pantheon+SH0ES, DESI DR II BAO +CMB, and $H(z)$ + Pantheon+SH0ES + DESI DR II BAO + CMB to solve the modified Friedmann equations and construct a Hubble parameter model whose free parameters are constrained through using Bayesian statistics through MCMC analysis. The observational analysis results in well-constrained free model parameters values. Specifically, the estimated $H_0$ values, derived from the different datasets combinations is found to be consistent with current cosmological probes, while the barotropic EoS parameter ($\omega$), lies in the physically viable quintessence regime. The constrained values serve as the basis for examining the cosmological evolution predicted by the suggested model. We investigated a number of significant cosmological quantities using these constrained values of the free parameters. The deceleration parameter exhibits a smooth evolution from an early epoch dominated by decelerated expansion to the present accelerating universe. Furthermore, the effective EoS and the effective DE EoS remain entirely in the negative region and approach the quintessence regime at the present and late-times, illustrating the present model provides a physically consistent description of the observed cosmological evolution. The behaviors of the viscosity pressure and effective pressure are also investigated. It is observed that, for all three combinations of observational datasets, both quantities stay negative over the considered redshift intervals. The persistent negative contribution of these pressures demonstrates the crucial role of viscosity in the cosmic dynamics and supports the accelerated expansion predicted by the model. Finally, a comparison of the constrained values of $H_0$ with different early- and late-universe measurements describes that the estimates obtained from the present analysis are compatible with a wide range of published observational results within their corresponding uncertainties.\\

Overall, the current study demonstrates that a viable and self-consistent cosmological framework can potentially be obtained by introducing viscous fluid with barotropic EoS to $f(T,\mathcal{L}_m)$ gravity. The model successfully reproduces the observed background expansion history, develops physically acceptable cosmological parameters, and is consistent with current observational constraints. These results imply that viscous $f(T,\mathcal{L}_m)$ gravity offers an effective way to examine the late-era cosmic dynamics and motivate more research on cosmological perturbations. structure formation, and forthcoming high-precision observational datasets.

\section{References}
\bibliographystyle{utphys.bst}
\bibliography{references}

\end{document}